\documentstyle[12pt]{article}    
\def\frac #1#2{{{  #1\over {#2} }}}

\def\1{{{\bf 1}}}

\def\A{{{\cal A}}}
\def\C{{{\cal C}}}
\def\H{{{\cal H}}}
\def\B{{{\cal B}(\H)}} 
\def\F{{{\cal F}}}

\def\M{{\cal M}}

\def\r{{{\varrho}}}
\def\p{{{1 \over 2p}}}

\def\F{{\cal F}}

\def\M{{\bf M}}

\def\R{{\Bbb  R}}
\def\rh{{\varrho}}
\def\rhp{{{ \varrho^{1 \over p }}}}
\def\rh2p{{{ \varrho^{1 \over 2p}}}}
\def\rhm{{{ \varrho^{-{1 \over 2p}}}}}
\def\erp{{{R^{1 \over p }}}}
\def\erp2{{{R^{1 \over 2p }}}}

\def\R2m{{{R^{-{1 \over 2p} }}}}
\def\up1{{{ \frac{p-1}{2}}}}
\def\vp22{{{ \frac{p - 2}{2}}}}
\def\wp24{{{\frac{p - 2}{4}}}}
\def\zp44{{{\frac{p - 4}{4}}}}
\def\1{{{\bf 1}}}

\font\small=cmr7

\newcommand{\be}{\begin{equation}}
\newcommand{\ee}{\end{equation}}

\title{\LARGE{Implementability of Liouville evolution, Koopman and 
Banach-Lamperti theorems in classical and quantum dynamics}}
\author{ I. ANTONIOU$^{1,2}$, W. A. MAJEWSKI$^{3}$ and Z. SUCHANECKI$^{1,2,4}$\\
\\
\small ${}^1$ International Solvay Institutes for Physics and Chemistry\\  
\small
C.P. 231, Campus Plaine ULB\\ 
\small Bd. du Triomphe, 1050 Brussels, Belgium \\
\\ 
\small ${}^2$ Theoretische Natuurkunde\\ 
\small Free University of Brussels 
\\ \\ 
\small
${}^3$ Institute of Theoretical Physics and Astrophysics\\
\small University of Gda\'nsk\\ \\
\small
${}^4$ Institute of Mathematics \\
\small University of Opole \\ \\ 
}  

\date{}     

\begin{document}
\maketitle

\baselineskip18pt 

\thispagestyle{empty}

\begin{abstract}  We extend the concept of implementability of semigroups
of evolution  operators associated with dynamical systems to quantum case.  We
show that such an extension can be properly formulated in terms of Jordan
morphisms and isometries
on non-commutative $L^p$ spaces. We focus our attention on a non-commutative 
analog of the Banach-Lamperti theorem. 
\end{abstract}

\vspace{1 cm}

\section{Introduction}

Operator theory and the associated theory of semigroups proved to be one of the
most successful methods elaborated for the study of dynamical systems. The idea
of using operator theory is due to  Koopman who replaced the time evolution
$S_t$ of single points from a phase space $\Omega$  by the time evolution of the
corresponding Koopman operators $V_t$ defined as
$$  
V_tf(\omega)=f(S_t\omega)\,,\ \ \ \ \ \ \ f\in L^2(\Omega),\ \omega\in \Omega\,.
$$  

Koopman \cite{Ko} introduced these operators in 1931 in order to study the
ergodic properties of dynamical systems using the powerful tools of operator
theory. This approach has been extensively used thereafter in statistical
mechanics and ergodic theory. The objects under consideration are Koopman
operators regarded as operators on $L^p$ spaces, $1\le p\le\infty$, and their
adjoints called Frobenius-Perron operators. Frobenius-Perron operators
describe, in particular, the evolution of probability densities defined on the
phase space $\Omega$. 

 The application of operator theory to dynamical
systems simplifies the study of their ergodic properties such as ergodicity,
mixing and exactness, as well as Kolmogorov systems, which is the
basis of the modern theory of
chaos \cite{KSF,LM}. Particularly important is the spectral analysis of
evolution operators that enables to extract important information about
their dynamical properties such as for example, the rate of the convergence to equilibrium.
Recent results obtained by the Brussels group (see
\cite{ASinf} and the references therein) show  that for unstable dynamical
systems there exist spectral decompositions of the evolution operators in terms
of resonances and resonance states, which appear as eigenvalues and
eigenprojections of the evolution operators. Another powerful method for the
study of unstable dynamical systems is based on the concept of a time operator
\cite{Mi,Pr}, which is defined as a selfadjoint operator $T$ associated with 
the evolution semigroup $V_t$ through  
the commutation relation 
$$  
TV_t=V_tT+tV_t\,.
$$  
The dynamical systems that admit time operators are highly unstable like 
Kolmogorov or exact systems. Nevertheless, the knowledge of the eigenvectors of
$T$ amounts to a probabilistic solution of the prediction problem for the
dynamical system described by the semigroup $\{V_t\}$. 

The time operator method serves to elucidate the problem of irreversibility in
statistical physics which is related to the understanding of the relation 
between reversible dynamical laws and the observed entropy increasing
evolutions.  Misra, Prigogine and Courbage \cite{MPC} showed that the unitary
evolution $U_t$ of a Kolmogorov system
can be intertwined with a Markov semigroup $W_t$, $t \ge 0$, through a
non-unitary transformation
$\Lambda$:
\be
 W_t \Lambda = \Lambda U_t \ , \quad t \ge 0 \,.
\label{intertw}
\ee
The intertwining transformation $\Lambda$ in the Misra-Prigogine-Courbage
approach is a non-increasing function of the time operator. 

The evolution operators that arise from point transformations of a phase space
are often modified, like in the  Misra-Prigogine-Courbage theory of
irreversibility, leading to new evolution semigroups that need not to be
related with the underlying point dynamics. The natural question is: Are such
operators associated with other point transformations? In other words, we ask
if, for example, modifications made on the level of evolution operators
correspond to some modifications on the level of trajectories in the phase
space. In a more general setting, we ask which linear operators on $L^p$ spaces
are implementable by point transformations?
A related question is: Which time evolutions of states of physical 
systems that are described in terms of a semigroup $\{W_t\}$ of maps on an 
$L^p$-space can be induced by Hamiltonian flows? There are some partial answers
to the above questions that will be presented below. 

In quantum mechanics we can also distinguish two levels of evolution of states
and observables that can be expressed in terms of evolution operators and 
semigroups. Thus
similar questions, as in the classical case, concerning implementability can
be raised. The quantum case is however more complex both technically and
conceptually. The evolution operators act on non-commutative $L^p$ spaces that
have much more complex structure and require sophisticated tools from operator
algebras theory. Secondly the very basic concept of implementability requires
clarification. 

In this article we formulate the  implementability in quantum case and prove
analogs of some classical results. The paper is organized as follows. In
Section 2 we give an overview of the results on implementability in classical
case. Section 3 contains an introduction to non-commutative $L^p$ spaces and
quantum dynamics. The formulation of quantum implementability and our main 
results are in Section 4.

\section{Classical case}
Let $(\Omega,\Sigma,\mu)$ be a measure space with a finite measure $\mu$.
A one parameter evolution semigroup $\{S_t\}$  of measurable transformations of
the space $\Omega$ defines a dynamical system. The variable $t$ 
signifies time and is continuous for flows and discrete for cascades. For
reversible systems $\{S_t\}$ is a group of automorphisms of $\Omega$. The space
$\Omega$ equipped with a measure structure is called the phase space and
measure $\mu$ represents in the case $\mu(\Omega)=1$ an equilibrium
distribution. 

The phase functions $f$  evolve according to the Koopman operators
\be
 V_tf(\omega)=f(S_t\omega) \ , \ \ \omega\in\Omega \,. 
\label{1}
\ee  
The Koopman operators are isometries on the Banach space 
$L^p=L^p(\Omega, \Sigma,\mu)$,  $p\ge1$, of $p$-integrable functions 
provided that $\{ S_t \}$ are measure preserving transformations.
If $S_t$  are automorphisms the Koopman operators restricted to the
Hilbert space $L^2$ are unitary. 

Consider the case of discrete time $t=1,2,\dots$ when the evolution
semigroup $\{V_t\}$ is determined by a single transformation $S$
$$
V_n=V^n\ \ \ {\rm and}\ \ Vf(\omega)=f(S\omega)\,.
$$

The relation of the point dynamics with the Koopman operators is clarified by
asking the question: What types of isometries on   $L^p$
spaces are implementable by point transformations? For $L^p$ spaces with
$p\not=2$, all isometries induce underlying point transformations.
 Such theorems on the implementability of isometries on   $L^p$ spaces, $p\neq
2$, are known as Banach-Lamperti theorems \cite{Ba,La}. The converse to 
Koopman's lemma in the case $p=2$, which holds under the additional assumption
that the isometry on $L^2$ is positivity preserving, can be found in
\cite{GGM}. The result is that an isometry $V$ is implementable by a necessarily
measure preserving transformation $S$
$$ Vf(\omega)=f(S\omega) \ , \ \ \omega\in\Omega \,. 
$$ 

The just quoted results on the relations between the point dynamics and
Koopman's maps on $L^p$-spaces gain additional interest if we realize that
$L^p$-structures can be used in the analysis of  a very large class of
dynamical systems. To describe briefly how $L^p$-techniques may be used for
such analysis let us assume that the state space $\Omega$ is a compact metric
space and 
$\Sigma$ is the $\sigma$-algebra   of Borel sets.
 Let us consider a (homogeneous) Markov
process on $\Omega$.
Then, the Kopmman's-like construction leads to the well defined
semigroup $W_t$ on the set $L^{\infty}(\Omega)$ of all bounded measurable
functions on $\Omega$. Let us restrict ourselves to Markov-Feller processes,
i.e. such processes that $W_t : C(\Omega) \to C(\Omega)$, where 
$C(\Omega)$ denotes the set of all continuous complex-valued
functions on $\Omega$. The role of the set $C(\Omega)$ for the quantization
procedure will be explained in the next section. Then, one
can show that any Markov-Feller process induces a positivity preserving
semigroup on $C(\Omega)$ and, conversely, with each such a semigroup is
associated a Markov-Feller process. Moreover,   such semigroups on $C(\Omega)$
can be extended to a semigroup  of contractions on $L^p(\Omega,\mu)$, where 
$p \ge 1$ and a measure $\mu$ is time invariant  (here, $\mu$ is called a time 
invariant measure if $\mu(W_tf) = \mu(f)$, $f \in C(\Omega)$). We recall that
Markov-Feller processes constitute an important tool in the description of real
systems of interacting  particles (see \cite{Li}). Consequently, we can study
stochastic processes in real physical models in terms  of semigroups on
$L^p$-spaces. The important point to note here is that such an approach yields 
the  possibility of studying, in an effective way, various ergodic  properties
of the considered processes, e.g. the question of convergence to  equilibrium,
question of spectral gaps, hypercontractivity (i.e.,  a set of ideas in field
theory, important for determination of the best constants in classical
inequalities and bounds on semigroup kernels, cf. \cite{DGS}), and finally to
utilize  various types of inequalities (e.g. log Sobolev and Nash inequalities,
see \cite{Gros}  as  well as \cite{DGS} and references therein).

Before addressing the question of implementability of Misra-Prigogine-Courbage
semigroups introduced in the previous section let us first recall some basic
facts.  Consider an abstract dynamical flow given by the quadruple $(\Omega
,\Sigma,
 \mu, \{S_t \})$ , where  $\{S_t \}$ is a group of one-to-one $\mu$ invariant
transformations of $\Omega$ and either $t\in {\bf Z}$ or $t\in {\bf R}$.
 The invariance of the measure $\mu$ implies
that the transformations $U_t$ 
$$
 U_t\,\rho(\omega) = \rho (S_{-t}\omega), \quad \rho \in L^2
$$
 are unitary operators 
on
$L^2$. Generally, $U_t$ is an isometry on the space $L^p$, $1 \le p \le \infty$.
Let us point out the following, very important properties of $U_t$ as 
operators on $L^1$:
\begin{itemize}
\item[(a)] $U_t \rho \ge 0$ if $\rho \ge 0$,
\item[(b)] $\int_\Omega U_t \rho \, d\mu = \int_\Omega \rho\, d\mu$\,, for $\rho
\ge 0$,
\item[(c)] $U_t1 = 1$. 
\end{itemize}

An abstract operator $W$ on $L^1$ which satisfies conditions (a)--(c) is
called {\it doubly stochastic operator}.

The Misra-Prigogine-Courbage theory of irreversibility \cite{MPC} (see \cite{Su}
for its generalized version) proposes to relate the group  $\{U_t \}$,
considered on the space $L^1$, to the irreversible semigroup $W_t$, $t\geq 0$,
through a nonunitary, doubly stochastic operator $\Lambda$: 
$$
W_t\Lambda  = \Lambda U_t\,,\ \ t\geq 0\,,
$$
The operators $W_t$, which also form a doubly stochastic semigroup on $L^1$,
should tend strongly to the equilibrium state, as $t\rightarrow \infty$,
on some subset of admissible densities.
A dynamical system for which such a construction is possible is called {\it
intrinsically random} and the conversion of the reversible group  $\{U_t \}$ 
into the irreversible semigroup $\{W_t \}$ through a nonunitary transformation
$\Lambda$ is called a {\it change of representation}.

So far all known constructions of the operator $\Lambda$ have been done for 
dynamical systems which are K-flows. Let us recall that a dynamical system is
a K-flow if there exists a sub-$\sigma$-algebra
${\Sigma}_0$ of ${\Sigma}$ such that for ${\Sigma}_t = S_t({\Sigma}_0)$ we have 

\begin{itemize}
\item[(i)] ${\Sigma}_s \subset {\Sigma}_t$, for $s < t$
\item[(ii)] $\sigma\bigl( \cup_{t\in {\bf R}} {\Sigma}_t) = {\Sigma}$
\item[(iii)] $\cap_{t\in {\bf R}} {\Sigma}_t = {\Sigma}_{-\infty}$ -- the
trivial $\sigma$-algebra, i.e. the algebra of sets of measure 0 or 1
\end{itemize}
where $\sigma\bigl( \cup_{t\in {\bf R}} {\Sigma}_t)$ stands for
$\sigma$-algebra generated by 
${\Sigma}_t$, $t \in {\bf R}$.
The main idea of the construction of $\Lambda$ is the
following. With any K-flow we can associate a family of conditional
expectations $\{E_t\}$ with respect to the $\sigma$-algebras  $\{\Sigma_t\}$
(projectors if we confine ourselves to the Hilbert space  $L^2$).  These
projectors determine the time operator $T$: 
\be
T = \int_{-\infty}^{+\infty} t dE_t \ .
\label{to}
\ee 
Then $\Lambda$ is defined, up to constants, as a function of the
operator $T$:
\be
\Lambda = f(T) + E_{-\infty} \,,
\label{lam}
\ee
 where $E_{-\infty}$ is the
expectation (projection on constants).
The function $f$ is assumed to be positive, non increasing,
$f(-\infty) = 1$, $f(+\infty) = 0$ and such that $ln f$ is concave on {\bf R}.

The Markov operators $W_t$ are of the form
\be
W_t=\left(\int_{-\infty}^{\infty}{f(s)\over
f(s-t)}\,dE_s+E_{-\infty}\right)U_t\,.
\label{wt}
\ee
 
It should be clear that each operator $U_t$ is the Frobenius-Perron operator
associated with $S_t$ and thus it is the adjoint of the corresponding Koopman operator.
The operators $W_t$ preserve the property of double stochasticity
characteristic to Frobenius-Perron operators. Therefore the question is: are
$W_t$ Frobenius-Perron operators associated with some measure preserving
transformations $\tilde S_t$ or, equivalently, is the adjoint $W^*_t$ the
Koopman operator 
$$
W^*_tf(\omega)=f(\tilde S_t\omega)\,.
$$
As we have shown in Ref. \cite{SAT} the answer to this question is in general
negative. Only the choice of $\Lambda$ as a coarse graining projection gives
implementability \cite{AG}.

\section{Quantum case - non-commutative $L^p$-spaces}

Passing to quantum theory it is convenient to rewrite the previously described 
scheme in terms of quantum ``phase'' space. The underlying  philosophy is based
on the general observation that various categories of spaces, in particular the
classical  phase space $\Omega$, can be completely described by the
(commutative) algebras of functions on them (the phase space $\Omega$ by the
algebra of continuous functions
$C(\Omega)$). The idea of (algebraic) quantization then is that the 
corresponding non-commutative algebra ($C^*$-algebra $\A$) may be viewed as an
algebra of functions on a virtual ``non-commutative space'' (``quantum phase
space'' in our case). Such approach has proven to be very powerful in
contemporary mathematics:  for instance
the analysis of the algebra of all continuous functions on a  
topological group led to the notion of
quantum group. Moreover, this approach 
is a starting point for studying ``geometrical
properties'' of non-commutative algebras (cf. \cite{KV,JC}).

Within that scheme, we are able to discuss the relation between point dynamics
and Koopman's operators  for the Quantum Mechanics setting.
Namely, according to the above strategy point dynamics may be viewed as a
one parameter family of maps $\hat{S}_t$ on a $C^*$-algebra $\A$. Clearly,  in
this way we also include general  quantum Markov-Feller dynamics
into the considered framework for quantizing dynamical systems. Further, the Koopman's maps
$\hat{V}_t$ will be defined on non-commutative  $L^p(\A)$ spaces which are 
quantum analogues of classical $L^p(\Omega, \Sigma, \mu)$ spaces.
The relationship between $\hat{S}_t$ and $\hat{V}_t$ expresses the 
implementability of Liouville evolution for quantum systems. 

To implement the just given programme we start 
with the algebraic reformulation of the theory of quantum dynamical systems. 
To this end we note that observables
in quantum mechanics are described by selfadjoint operators on some
Hilbert space
${\cal H}$ and physical states by positive tracial operators on ${\cal H}$. In
the mathematical formalism  it is more convenient to consider a von Neumann
algebra ${\cal M}$ as the algebra of observables and its dual ${\cal M}^*$ as
the algebra of states. Thus, in the algebraic reformulation of the classical 
dynamical system
$(\Omega,\Sigma,\mu,\{ S_t\})$ we consider the
commutative $W^*$-algebra ${\cal A}= L^{\infty}(\Omega,\Sigma,\mu)$. 
Then, the
semigroup of Koopman operators will be replaced by the semigroup of 
homomorphisms of
${\cal A}$ which are given by the formula
$$ [\alpha_t(a)](f)(\omega)=a(S_t\omega)f(\omega)\,,
$$ 
for each function $a\in{\cal A}\subset {\cal B}(L^2)$ (i.e. $a$ is treated as
a bounded operator on
$L^2$) and $f \in L^2$. 

Turning to the quantum case, let us consider as a non-commutative analog of a 
probability space
$(\Omega ,\Sigma,\mu )$ the triple  $ (B({\cal H}), {\cal H}, \varrho) $
where $\cal H$ is a separable Hilbert space, $B({\cal H})$ is the set of all
linear bounded operators on $\cal H$ and $\varrho$ is a density matrix. Let us
assume that $\varrho$ is an invertible operator, which implies that 
$\omega(\cdot) = Tr \{ \varrho\, \cdot \}$ 
is a faithful state on $B({\cal H})$. 
In physical terms, $\varrho$ can represent, for example, a Gibbs state at a
temperature $\beta$.  Suppose that the dynamics of the system is given in the
Heisenberg picture,  i.e. the time evolution of the system is  given by a
one-parameter family of maps 
 $\alpha_t : B({\cal H})
\to  B({\cal H})$. More precisely, the equivalence of the Schroedinger 
and Heisenberg picture for reversible dynamics says, \cite{Kad}, 
 that the dynamics of observables 
can be given by a one parameter group $\alpha_t$ of Jordan automorphisms 
(that is linear, $*$-preserving, one-to-one and onto maps defined on a 
$C^*$-algebra  $\C$ such that
$\alpha_t(A^2) = \alpha_t(A)^2$ for $A \in \C$).

Treating $( \B, \H ,\r, \alpha_t)$ as a quantum analogue of a classical 
quadruple $(\Omega, \Sigma, \mu, S_t)$ we will introduce basic examples  of
quantum 
$L^p$-spaces as follows.
Observe first that the set of all
Hilbert-Schmidt operators
$\F_{H-S}$, on the space $\H$, has a Hilbert space structure with the inner
product given by 
$$(a,b)= Tr\{ a^*b \}\,,\ \ a,b \in \F_{H-S}\,.
$$ 
Therefore $\F_{H-S}$ can be considered as the quantum $L^2$-space associated
with  ``the quantum uniform measure'':
\be
 \omega_0(\cdot) \equiv Tr\{ {\bf 1} \cdot \}\,.
\label{qm}
\ee
Analogously, the set of all tracial operators (density matrices) ${\cal
F}_T$ can be regarded as the quantum $L^1$-space associated with the measure
(\ref{qm}). The corresponding norms for these spaces are:
 $$
|| \cdot||_p \equiv \bigl(Tr| \cdot |^p 
\bigr)^{1 \over p}\,,\ \ p=1,2\,.
$$ 

 Let us generalize the idea of quantum $L^p$ spaces 
to an arbitrary  ``quantum measure'' $\omega(\cdot) \equiv Tr\{\varrho\, \cdot
\}$ on $\B$. Let us fix $A_0 \ge 0$ and put 
$$
A_{\r} \equiv \rh2p A_0 \rh2p\,,
$$ 
where $p=1,2$. 
Observe that
$$
A_{\r} \le ||A_0|| \rhp
$$ 
and
$$ 
Tr A_{\r}^p = Tr A_{\r}^{\up1} A_{\r} A_{\r}^{\up1} \le ||A_0||
Tr \{ A_{\r}^{\up1} \rhp A_{\r}^{\up1} \}
\le ||A_0||^{p} Tr \varrho \le \infty\,.
$$
The notation was chosen in such a way that
it is easy to generalize the result to other $p's$.
Let us note that for an arbitrary operator $A\in \B$ 
$$
A = {{A + A^*} \over 2} +
i {{A - A^*} \over 2i} \equiv A_h + iA_a
\equiv A_h^+ - A_h^- + iA_a^+ - iA_a^-\,,
$$ 
where $A_h^+, A_h^-, A_a^+, A_a^-$
are positive operators.
Therefore
$$
\begin{array}{ll}
\bigl(Tr |\r^{\p}A \r^{\p}|^p \bigr)^{1 \over p}&\le
\bigl(Tr |\r^{\p}A_h^+\r^{\p}|^p \bigr)^{1 \over p} + \bigl(Tr |\r^{\p}A_h^-
\r^{\p}|^p \bigr)^{1\over p}\\
 &
 +\bigl(Tr |\r^{\p}A_a^+ \r^{\p}|^p \bigr)^{1 \over p}
+\bigl(Tr |\r^{\p}A_a^- \r^{\p}|^p \bigr)^{1 \over p}\\ 
&
< \infty\,,
\end{array}
$$
which implies that we can relate to each $A\in\B$ a trace class operator:
\be
\B \ni A \mapsto |\r^{\p}A \r^{\p}|^p \in {\cal F}_T\,, 
\label{2}
\ee
for $p \in \{1,2\}$, and for any fixed (arbitrary) density
matrix $\r$. Consequently, it is easy to see that
\be
||A||_p \equiv \bigl( Tr |\r^{\p}A \r^{\p}|^p \bigr)^{1 \over p}
\label{3}
\ee
is a well defined norm on $\B$.
The Banach space defined as the completion of $\B$ in the norm
(\ref{3}) will be denoted as $L^p(\B,\omega)$. As expected, the $L^2(\B,
\omega)$ is a Hilbert space with the scalar product
\be
<A,B> \equiv Tr(\r^{1 \over 2}A^* \r^{1 \over 2} B)
\label{4}
\ee

The above result saying that (\ref{3}) is a well defined norm
on $\B$ can be easily extended to an arbitrary $p \ge 1$
(see \cite{Tr,Zo}, and \cite{MZ2} for details). Moreover, it can
be shown that the spaces  $L^p(\B, \omega)$
and $L^q(\B, \omega)$, with $p,q \in (1, \infty)$,
${1 \over p} + { 1 \over q} = 1$, are duals of each other. 
One can also introduce the space
$L^{\infty}(\B, \omega)$ as the dual to $L^1(\B, \omega)$.
Such a construction ensures that we have
\be
L_p(\B, \omega) \subseteq L_q(\B, \omega)
\label{5}
\ee
for $1 \le p \le q \le \infty$. Finally, note that
the one parameter family of Banach spaces $\{ L^p(\B, \omega) \}_{p \ge 1}$ 
forms a so-called interpolating scale, i.e. the Banach spaces $L^p(\B, \omega)$
with $1 \le p \le \infty$ are interpolating spaces between $L^1(\B, \omega)$
and $L^{\infty}(\B,\omega)$.
In particular, interpolation theory (like in the classical case) is also 
available in this case. Therefore, a large number of ``classical'' 
$L^p$-estimates is also  available for quantum dynamical systems (cf.
\cite{Ter}).
\smallskip

As an example, let us reconsider a very special case
in the above construction. Namely, instead of the state $\omega$
let us take $\omega_0 \equiv Tr\{ {\bf 1} \cdot \}$.
In mathematical terms we replace state by the weight $Tr(\cdot)$.
Then, the repetition of the above argument leads to
\be
 {L}^p(\B, Tr) \equiv \{ A \in \B; ||A||_p 
= (Tr |A|^p)^{1 \over p} < \infty \}.
\label{6}
\ee
It is easy to recognize that we get all $p$-Schatten classes.
In other words, the trace class operators as well as
the Hilbert-Schmidt operators, mentioned at the beginning
of this section, constitute special cases of quantum 
$L^p$-spaces.
We recall that these spaces have been used for the study of
various problems of quantum statistical physics and that 
such an approach is called the quantum Liouville space
technique (see \cite{Em}). For a slightly different definition of
$L^2$-spaces associated with a quantum state and their
applications to ``probabilistic'' descriptions of quantum systems
see Chapter II in \cite{Ho}.
\smallskip

Before proceeding with the construction of non-commutative Koopman's 
operators let us summarize here various general $L^p$-spaces 
that we will need in the next section. We start with the observation that
a general formulation of a quantum schema deals with
a general $W^*$ (or even $C^*$) algebra $\A$ describing individual
properties of a fixed physical system (see \cite{BR,Haa,Ru}). 
In particular, the basic procedure of statistical
mechanics, the thermodynamic limit, leads to a $C^*$-algebra
$\A$ which can be very different from that of $\B$.
Consequently, general quantum $L^p$ spaces 
corresponding to a quantum system should be based on a general $C^*$-algebra.
However, such a general construction of
non-commutative $L^p$-spaces  is rather involved
(see \cite{MZ2,Ter,Haa,Kos,MZ1,MZ3}).
Nevertheless,
such a general scheme has proved to be very useful
for the description of concrete models with quantum Markov dynamics
(see \cite{MZ2,MZ1,MZ3,AH,Ci,DL}).

As it is not our purpose to study here the mathematical
questions related to a construction of general $L^p$-spaces
we shall restrict
ourselves to the case of a von Neumann algebra $\M$ with a 
 faithful semifinite normal ({\it fsn}) trace $\varphi$.
Clearly,  $ \B$ has this property, i.e the case of Dirac's quantum
mechanics will be included. Therefore, we can consider the pair  $\{ \M,
\varphi \}$  consisting of a von Neumann algebra and 
{\it fsn} trace. 
Let $\omega$ be a normal linear functional on $\M$.
Then (see \cite{Se}) $\omega$ is of the form
$\omega(a) = \varphi({ R}a)$, $ a \in \M$, 
where $R$ is an $L^1$-integrable (so 
$\varphi(|R|) < \infty$)
uniquely determined non-singular positive operator.
Define
\be
||A||_p \equiv \bigl( \varphi |{R}^{\p}A {R}^{\p}|^p 
\bigr)^{1 \over p}
\label{3a}
\ee
It can be proved that $|| \cdot ||_p$, $p \ge 1$, is a norm on $\M$
(see \cite{Tr,Zo}). The completion of $\M$ with respect to this norm
again leads to the non-commutative $L^p(\M, \omega)$
Banach space (see \cite{Tr,Zo,Se,Dix}) with all the listed properties
of $L^p(\B, \omega)$-spaces.
Consequently, we got non-commutative $L^p$-spaces which can be associated 
with a large family of quantum dynamical systems.

\section{The converse of Quantum Banach-Lamperti 
theorem}

We have seen that the Koopman's construction (\ref{1})
gives a well defined {\it bounded} map $V_t$ on $L^p$-spaces,
so $V_t$ can be considered as an ``integrable'' map.
Our first observation is that we have analogous 
situation in the non-commutative 
framework. Namely, let us assume that $T: \M \to \M$ is a
linear bounded map. Here and subsequently, $(\M,\omega)$ denotes a semifinite 
von Neumann algebra
and a state on it respectively.
Denote by $\iota_p$ the imbedding of $\M$ into 
$L^p(\M, \omega)$ and define operator $T^{(p)}: L^p \to L^p$ by formula:
\be
 T^{(p)}(\iota_p(a)) = \iota_p(Ta) \quad a \in \M
\label{m7}
\ee
We say that $T$ is $p$-integrable (with respect to $\omega$)
if the induced operator $T^{(p)}$ is $L^p$-bounded, in which case we denote
its unique extension to $L^p(\M, \omega)$ by the same letter.
We have the following useful, and in fact very general,
criterion for quantum integrability
(see \cite{GL}). Let $T: \M \to \M$ be a normal, positivity preserving
linear map. Then $T$ is integrable with respect to $\omega$ if and only if
$T_*\omega \le {\rm Const} \circ \omega$ where $T_*$ denotes the predual
map. We want to add that we dropped here the ``$p$th'' in the word integrability
as one can show that $p$-integrability implies $r$-integrability
for $r \ge p$ (see \cite{GL}).
In particular, if $J: \M \to \M$ is a Jordan
automorphism  satisfying
$\omega \circ J \equiv \omega$ then $J^{(p)}$ is an integrable map. 
One can even show that $J^{(p)}$ is an isometry.

Having clarified this point let us turn to a quantum analogue of
Banach-Lamperti result (another approach
to that question was recently given in Ref. \cite{L}). 
Let  $\{ \M, \varphi \}$ be a von Neumann algebra with 
{\it fsn} trace and let $L^p(\M, \varphi)$, $p \ge 1$, be the corresponding
quantum $L^p$-space. Assume that $T$, $T: L^p(\M, \varphi) \to
L^p(\M, \varphi)$  is a linear map.
Then, (see \cite{Ye}, and \cite{Wat} for the most general case),
$T$ is $L^p$-isometry of $L^p(\M, \varphi)$ {\it onto}
itself if and only if 
\be
T(x) = W B J(x), \quad x \in  L^p(\M, \varphi) \cap \M
\label{8a}
\ee
where $W \in \M$ is unitary, $B$ a selfadjoint operator
affiliated with the center of $\M$ and $J$ a normal
Jordan isomorphism mapping $\M$ onto itself, such that
\be
\varphi(X) = \varphi(B^p J(X)),\quad {\rm for
\quad all \quad} X \in \M, \quad X \ge 0.
\label{8b}
\ee
Formula (\ref{8a}) is nothing but the statement that any $L^p$-isometry $T$
of $L^p(\M, \varphi)$ onto itself is implemented by a Jordan morphism $J$
which is multiplied by operators $W$ and $B$.

\vskip.5cm

In physics, especially in statistical physics, we are 
usually interested in $L^p$-spaces associated with a finite measure.
In other words we want to associate the $L^p$-space with 
the pair $(\M, \omega)$ where $\omega$
is a state on $\M$. To this end let us define the following map
\be
 \M \ni X \mapsto {R}^{\p}X{R}^{\p}
\label{m9}
\ee
where $R$ is the operator determined by
the equality $\omega(X) = \varphi (RX)$ (see also the
argument leading to formula (\ref{3a}).
One can show (cf \cite{Tr}) that (\ref{m9}) can be extended to an isometric
isomorphism $\tau_p$ between $L^p(\M, \omega)$ and $L^p(\M, \varphi)$.
Moreover, let $V$ be an isometry from $L^p(\M, \omega)$
onto $L^p(\M, \omega)$. Then, there exists an isometry
$T$ on $L^p(\M, \varphi)$ such that
the following diagram is commutative
$$
\matrix{ L^p(\M, \varphi) &  {\mathop{\longrightarrow}\limits^{T}}   &  
L^p(\M, \varphi) \cr
\uparrow \tau_p            &          &   \uparrow \tau_p\cr
L^p(\M, \omega)           & {\mathop{\longrightarrow}\limits^{V}}   &   
L^p(\M, \omega)\cr}
$$
Consequently, any isometry from $L^p(\M, \omega)$ onto $L^p(\M, \omega)$
has the following form:
\be
V(X) = \R2m T(\erp2 X \erp2) \R2m
\label{m11}
\ee

Now, let us restrict ourselves to the setting of Dirac's Quantum Mechanics.
So, we put $\M \equiv \B$ and consider a linear map 
$V: L^p(\B, \omega) \to L^p(\B, \omega)$
where $\omega(\cdot) = Tr\{ \varrho\, \cdot\}$.
We assume that $V$ is an isometry
of $L^p(\B, \omega)$ onto itself such that
$V(\1) = \1$ and $V(X) \ge 0 $, for $X \ge 0$. Then
\be
V(X) =  \rhm WB J(\rh2p X \rh2p) \rhm
=\rhm J(\rh2p X \rh2p) \rhm = J(X)\,, 
\label{m12}
\ee
for $X \in \B \cap L^p(\B, \omega)$,
where the second equality follows from the positivity
and identity preserving assumption as well as from the irreducibility of $\B$ 
while the third equality follows from the fact that each Jordan isomorphism can
be  split into the sum of $^*$-isomorphism and $^*$-anti-isomorphism.

\vskip 1cm
Having such a version of the converse of quantum Banach-Lamperti theorem
we are in a position to discuss the questions related to {\it a change of 
representation} (cf. Section 2) but, now, in a quantum mechanical setting.
Let the dynamical map $V_t : L^2(\B, \omega) \to  
L^2(\B, \omega)$ be induced by a hamiltonian flow. 
Can we change the representation,
that is to find a map $\Lambda : L^2(\B, \omega) \to L^2(\B, \omega)$ 
such that there is no information lost in the sense that
$\Lambda$ ( so also $\Lambda^{-1}$) is affine isomorphism
of the set of states $\cal S$ onto itself and with the property that
the composition : $\Lambda \circ V_t \circ \Lambda^{-1}$
no longer induced by a point dynamics?
To answer this question let us note that
the predual space $\A_*$ is isometric to $L^1(\B, \omega)$
(so we can identify them), 
$L^1(\B, \omega) \subset L^2(\B, \omega)$, ${\cal S}
\hookrightarrow L^1(\B, \omega)$, and
$\cal S$ is not subset of any hyperplane of $L^1(\B, \omega)$.
Then using Kadison's result (see \cite{Kad}) one can see
firstly that the affine isomorphism $\Lambda_0$ 
can be extended to a linear map $\Lambda$ 
on $L^1(\B, \omega)$. Secondly $\Lambda$ as a map
of $\B_*$ onto itself is induced by a Jordan automorphism
$\alpha: \B \to \B$. Finally using the quantum converse
to Koopman's theorem we infer that the composition
$\Lambda \circ V_t \circ \Lambda^{-1}$ is induced by the 
uniquely determined Jordan automorphism.
In other words, we obtained the negative answer to the above posed question.

The important point to note here is that we assumed $\Lambda$ to be 
an affine isomorphism while 
$V_t$ to be an isometry {\it onto}. To get the implementability of
$\Lambda \circ V_t \circ \Lambda^{-1}$ by a uniquely determined Jordan morphism
$\alpha: \B \to \B$ these conditions can not be weaken. In other words, relaxing the conditions
{\it ``isomorphism''} and {\it ``onto''} the implementability will be lost and again, we have
analogous situation to that in classical case. The question of the change of
representation through a non-isomorphic operator $\Lambda$ and the related 
question of implementability will be discussed in a forthcoming publication.

\section*{Acknowledgment} We thank L.E Labuschagne for helpful comments.  This
work enjoyed the financial support of the Belgian Government through the
Interuniversity Attraction Poles. This work has also been supported by KBN grant
PB/0273/PO3/99/16.

\end{document}